\documentclass[useAMS,usenatbib]{mn2e}
\usepackage{graphicx}

\title[The distance to the Leo I dwarf spheroidal galaxy]
{The distance to the Leo I dwarf spheroidal galaxy 
from the Red Giant Branch Tip
    \thanks{Based on observations made with the Italian Telescopio Nazionale
     Galileo (TNG) operated on the island of La Palma by the Centro
     Galileo Galilei of the INAF (Istituto Nazionale di Astrofisica) 
      at the Spanish Observatorio del Roque de los Muchachos
                  of the Instituto de Astrofisica de Canarias.
    Based on archive data from observations made with the ESO-VLT telescope, 
    at ESO, Paranal, Chile under programme 64.N-0421(A)}}
\author[M. Bellazzini et al.]{M. Bellazzini$^{1}$, 
N. Gennari$^{2}$, F.R. Ferraro$^{2}$, A. Sollima$^{1,2}$
\thanks{E-mail: bellazzini, ferraro@bo.astro.it}\\
$^{1}$INAF - Osservatorio Astronomico di Bologna, via Ranzani 1, 40127, Bologna,
Italy\\
$^{2}$Universit\`a di Bologna - Dipartimento di Astronomia, via Ranzani 1, 40127, Bologna,
Italy}

\begin{document}

\date{\today}

\pagerange{\pageref{firstpage}--\pageref{lastpage}} \pubyear{2003}

\maketitle

\label{firstpage}

\begin{abstract}
We present V and I photometry of a $9.4\arcmin \times 9.4\arcmin$ field 
centered
on the dwarf spheroidal galaxy Leo~I. The I magnitude of the tip of the Red
Giant Branch is robustly estimated from two different datasets ($I^{TRGB}=17.97
^{+0.05}_{-0.03}$). From this estimate, adopting  $[M/H]\simeq-1.2$ from the
comparison of RGB stars with Galactic templates, we obtain a distance modulus
$(m-M)_0=22.02 \pm 0.13$, corresponding to a distance $D=254^{+16}_{-19}$ Kpc.
\end{abstract}


\section{Introduction}

Leo~I is one of the brightest ($M_V=-11.9$) dwarf spheroidal galaxies of the 
Local Group and is the most distant Galactic satellite presently known 
\cite[$D=250\pm 30 $ Kpc, according to][]{mateo}.  Recent accurate studies
based on HST photometry \citep{caputo,carme} have firmly established that the
large majority of Leo~I  stars have an age comprised between 7-10 Gyr and 1
Gyr, but the dominance of intermediate-age population is known since the early
CCD analyses  \citep[see][for an overview of pre-HST
studies]{reid,lee_leo,gambu}.  The presence of a (sparse)  old population
(age$\ga 10$ Gyr) has been confirmed only very recently,  with the detection of
Blue Horizontal Branch stars \citep{held_hb} and RR Lyrae  variables
\citep{held_RR}. The available estimates of the mean metallicity range from
$[Fe/H]=-2.1$ to $[Fe/H]=-1.0$  \citep[see][for a review]{caputo}. The
uncertainties are large but there is a general consensus on the metal-poor 
nature of the dominant stellar population.

Given its large distance from the Milky Way and its large 
radial velocity, this
system has a crucial r\^ole in the estimate of the mass of the Galaxy at large
radii \citep{z89,z99,we99} and of the Local Group as a whole, via the {\em
Local Group Timing} technique \citep{lb99}. The uncertainty on its actual
distance is the major contributor to the error budget in this kind  of studies
\cite[see][for discussion and references]{z99,lb99}.

In this paper we provide a new estimate of the distance to the Leo~I 
galaxy using the Tip of the Red Giant Branch (TRGB) as a standard candle
\citep[see][for details and references about the method]{lfm93,smf96,mf98,scw}.
We obtain a robust estimate of the apparent magnitude of the TRGB in the I
passband and we derive the distance modulus of Leo~I adopting our  recent
calibration of the method \citep[][hereafter B04]{tip1,tip2} whose zero-point 
is fully
independent from the RR-Lyrae distance scale  \cite[that is still affected by
sizeable uncertainties, see][]{carla,walk}. The
uncertainty on the final distance modulus is estimated by mean of a Monte Carlo
simulation taking into account all the possible sources of error.
This work is part of a large programme to obtain homogeneous
distances for most of the galaxies of the Local Group \cite[see][]{draco,m33}.

The plan of the paper is the following: in Sect.~2 we describe the
observational material, the data reduction process, the artificial stars
experiments, and we briefy discuss the Color Magnitude Diagrams;
in Sect.~3 we report on the detection of the TRGB,
and on our estimate of the distance modulus. Finally, the main results 
 are summarized in Sect.~4.

\section{Observations, Data Reduction and Color Magnitude Diagrams}

The  data were obtained  at the  3.52 $m$  Italian telescope  TNG  (Telescopio 
Nazionale  Galileo - Roque  de  los  Muchachos, La  Palma, Canary Islands,
Spain),  using DoLoRes, a focal reducer imager/spectrograph   equipped with a
$2048 \times 2048$ pixels CCD array.  The pixel scale is $0.275$  arcsec/px,
thus the total  field of view of  the camera is $9.4\arcmin  \times 9.4\arcmin
$. The observations were carried out during  three nights (March 19, 20 and 21,
2001), under average seeing conditions  ($FWHM\simeq  1.0\arcsec -
1.4\arcsec$). The data have been acquired during the same observational run
already described in \citet{draco}: any further detail may be found in that
paper.

We acquired five 300s exposures in I, and two 600 s and three 300s  exposures
in V, centered on the center of Leo~I. All the  raw  images were  corrected 
for bias  and  flat  field, and  the overscan  region  was  trimmed  using
standard  IRAF\footnote{IRAF  is distributed by  the National  Optical
Astronomy Observatory,  which is operated by the Association of Universities
for Research in Astronomy, Inc.,   under  cooperative   agreement  with   the  
National  Science Fundation.} procedures. Each set of five images (per filter)
was registered,  flux-normalized and combined into one single {\em master
frame} using the  tasks {\em interp.csh} and {\em ref.csh} of the ISIS-2.1
package \citep{alard}. ISIS is able to combine images into a master frame
having the seeing of the best image in the set, without any loss of flux
\citep{alard}.

The PSF-fitting procedure was performed  independently on each V and I master
image, using a version of DoPhot \citep{doph} modified by   P.  Montegriffo at
the Bologna  Observatory to  read  images in  double precision  format. The
frames were searched for sources adopting a 5-$\sigma$ threshold, and the
spatial variations of the PSF were modeled with a quadratic polynomial. A 
final  catalogue listing  the instrumental  V,I magnitudes  for all  the  stars
in  each  field has  been obtained  by cross-correlating the V and  I
catalogues. Only the sources classified as  stars  by  the  code  have been 
retained. Aperture corrections have been determined on a sample of bright and
isolated stars in each of the master frames and applied to the catalogues.

The transformation to the standard Johnson-Cousins photometric system has been
achieved using the calibrating relation obtained and described in
\citet{draco}. The absolute calibration has been checked against independent
photometries \citep{stet,momany} and it has been found to be accurate at the
$\pm 0.02 / 0.03$ mag level \cite[see][]{draco}.

Leo~I is one of the best studied local dwarf spheroidals. Hence it is worth to
check if there are other available datasets that may provide support to the
present analysis. The search of the ESO archive has been successful. We
retrieved from the ESO archive one V and one I 90 s frames centered on Leo~I
taken with the FORS~1 imager at the VLT-Antu telescope, at Cerro Paranal,
Chile, on the night of December 2, 1999. The field of view is  $6.7\arcmin 
\times 6.7\arcmin $ and the pixel scale is $0.200$  arcsec/px. The images were
taken under excellent seeing conditions  ($FWHM\simeq  0.6\arcsec$). The images
were reduced in the same way described above and were calibrated directly on
the calibrated catalogues obtained from the TNG data. 
In the following we will refer
to the TNG and VLT dataset as to the {\em TNG sample} and {\em VLT sample},
respectively.

\begin{figure}
\includegraphics[width=84mm]{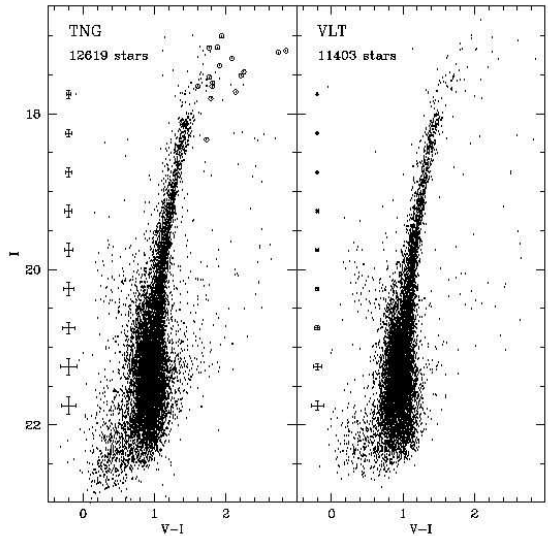} 
\caption{I,V-I CMDs of Leo~I obtained from TNG data (left panel) and VLT data
(right panel). The characteristic photometric uncertainties are plotted in the
two panels as errorbars aligned at $V-I=-0.2$. The open circles in the TNG CMD
mark the carbon stars we have identified from \citet{demers}.}
\end{figure}

\subsection{Color-Magnitude Diagrams}

The Color Magnitude Diagrams (CMD) of both the considered samples are displayed
in Fig.~1. The two diagrams are very similar but the higher accuracy of the VLT
relative photometry may be appreciated at a first glance, and it is probably
due to the combination of the larger collecting area of the telescope and the
much better seeing conditions. The morphology of the CMDs shown in Fig.~1 is
fully consistent with those presented and described in previous studies
\citep{reid,lee_leo,gambu,caputo,carme,held_hb}.

The CMDs are dominated by the steep RGB of Leo~I extending from the limiting
magnitude to $I\simeq 18.0$.  A handful of bright AGB stars 
\cite[see][]{lee_leo,agb} are also present above this luminosity,  in the range
$1.2\la V-I\la 3.0$. Among these stars we have marked with open circles (in the
TNG sample) the carbon stars identified by \citet{demers}. The compressed color
scale of the CMDs, that has been adopted to include the reddest stars at
$V-I\sim 3.0$,  prevents a clear view of the Red Clump (RC) of Helium-burning
stars that is quite evident in enlarged CMDs, peaking at $I \simeq 21.45$,
slightly bluer than the RGB sequence at that magnitude. The sparse plume of
stars around $I\sim 20$ and $0.0\la V-I\la 0.8$ has been identified by
\citet{caputo} as the high-mass tail of He-burning stars of Leo~I. 

\subsection{Artificial Stars Experiments}

To quantify the effects of the data reduction process on our photometry we
performed a set of artificial stars experiments on the TNG dataset.  We
followed exactly the procedure described in \citet{draco} and we refer the
interested reader to that paper for any detail. The artificial stars were
extracted from a LF similar to the observed one, with the additional
requirement that they must lie on the average ridge line representing the
observed RGB (see Fig.~2, upper left panel). We limited the artificial
stars experiments to RGB stars with $I\le 21.0$. 
The stars were added ($\sim 100$ at a time to avoid any spurious
modification of the actual crowding conditions) to the master frames and the
whole process of data reduction was repeated at any run. 

\begin{figure}
\includegraphics[width=84mm]{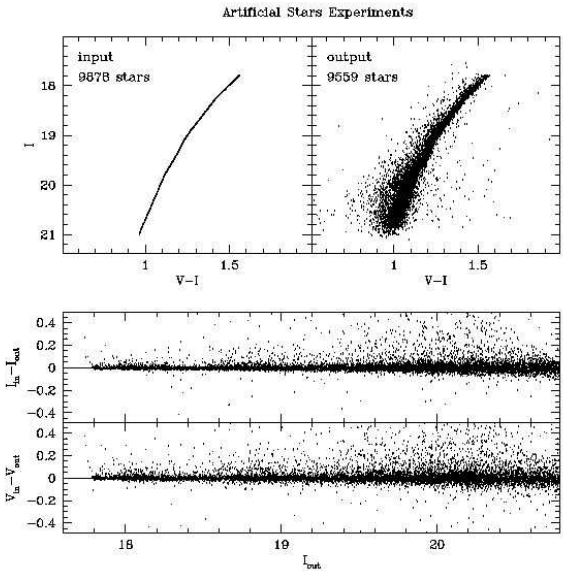} 
 \caption{Results of the artificial stars experiments. Upper left panel:
 CMD of the artificial stars in terms of {\em input} magnitude and color. 
 Upper right panel: CMD of the artificial stars in terms of {\em output} 
 magnitude and color. Lower panels: differences between input and output
 magnitudes as a function of $I_{out}$. The excess of stars in the half-part
 of the diagram of positive magnitude difference is due to blended stars that
 are recovered with as brighter sources with respect to their input 
 counterparts.}
\end{figure}

The difference between input and output magnitudes shown in the lower panels of
Fig.~2 confirms that the photometric errors are quite small and the degree of
blending cannot affect our analyses. The completeness of the sample (not shown)
is larger than 80\% in the whole range of magnitude considered by our
experiments.

\section{The distance to Leo~I}

\subsection{Detection of the TRGB} 

The use of Tip of the Red Giant Branch (TRGB) as a standard candle is now a
mature and widely used technique to estimate the distance to galaxies of any
morphological type  \cite[see][for a detailed description of the method, recent
reviews and applications]{lfm93,mf95,mf98,walk}. The underlying physics is well
understood \citep{mf98,scw} and the observational procedure is operationally
well defined \citep{mf95}. The key observable is the sharp cut-off occurring at
the bright end of the RGB Luminosity Function (LF) that can be easily detected
with the application of an edge-detector filter \citep[Sobel
filter,][]{mf95,smf96} or by other (generally parametric) techniques \cite[see,
for example][]{mendez,mcc}. The necessary condition for a safe application of
the technique is that the observed RGB LF should be well populated, with more
than $\sim 100$  stars within 1 mag from the TRGB \citep{mf95,draco}. 
This criterion is clearly fulfilled in the present analysis since there are 
$N_{\star}=346$ and $N_{\star}=298$ stars within one magnitude from the 
detected TRGB in the TNG and VLT samples, respectively,

The detection of the TRGB is displayed in Fig.~3 and 4 for the TNG sample and
the VLT sample, respectively. In both cases, the cut-off of the RGB LF is
clearly evident and it is easily detected by the Sobel's filter. We take the
position of the main peak of the filter response as our estimate of $I^{TRGB}$
and the Half Width at Half Maximum of the peak as the associated uncertainty
($I^{TRGB}=17.98^{+0.09}_{-0.06}$ for the TNG sample, and 
$I^{TRGB}=17.97^{+0.06}_{-0.03}$ for the VLT sample). The
higher photometric precision of the VLT sample allows a more accurate
location of the Tip, with respect to the TNG sample.  As our final
estimate we take the weighted mean of the two detections, 
$I^{TRGB}=17.97^{+0.05}_{-0.03}$. 

\begin{figure}
\includegraphics[width=84mm]{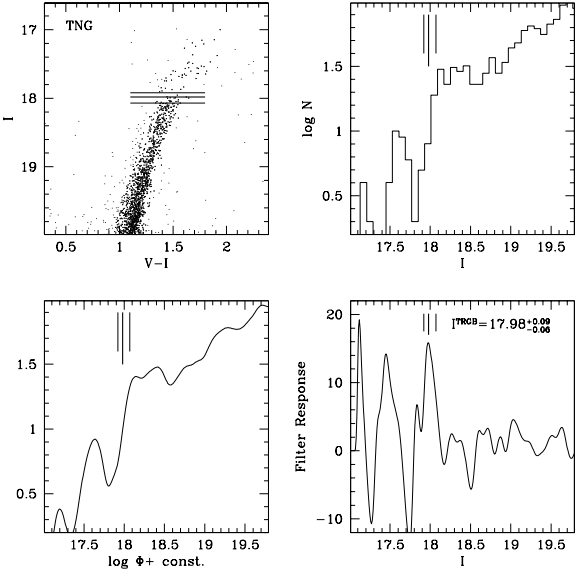} 
\caption{TRGB detection from the TNG sample. The stars selected to obtain the
RGB LF are plotted as heavy points in the CMD shown in the upper left panel.
The logarithmic LF of the upper portion of the RGB is shown as an ordinary 
histogram (upper right panel) and as a generalized histogram 
\citep[i.e., convolved with a gaussian, see][]{laird}. 
The lower right panel shows the Sobel's filter response to the LF. 
In all panels the thick line marks the position of the TRGB,
while the thin lines enclose the error bars of the estimate.}
\end{figure}

\begin{figure}
\includegraphics[width=84mm]{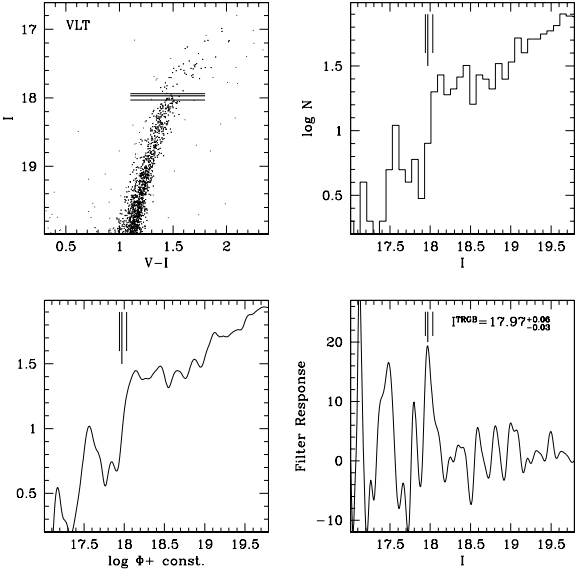} 
\caption{TRGB detection from the VLT sample. 
The symbols are the same as in Fig.~3.}
\end{figure}

Our estimate of $I^{TRGB}$ is in excellent agreement with the results by
\citet{momany} and \citep{caputo}. 
The estimate by  \citet[][$I^{TRGB}=18.25\pm 0.10$]{lee_leo} 
is $\sim 1.9 \sigma$ larger than our value. Part of the
difference may be due to insufficient sampling of the RGB LF, since their field
is significantly smaller than ours  ($5.7\arcmin \times 5.7\arcmin$, 2.6 times
smaller than the TNG field and 1.4 times smaller than the VLT field). This
conclusion is also supported by the comparison with \citet{mendez} which states
that in their   $5.5\arcmin \times 7.3\arcmin$ field they barely collect 100
RGB stars in the upper 1 mag from the TRGB. These authors find $I^{TRGB}=18.14
\pm 0.11$, within one $\sigma$ of our estimate, but still appreciably fainter.
Also in this case, problems of sampling and/or of absolute photometric
calibration may be at the origin of the difference with our result. We regard
our result as the most robust since (a) the sampling of the upper-RGB LF 
is excellent in both datasets and, (b) our photometric calibration has been 
successfully compared with various independent photometries.

\subsection{The distance modulus of Leo~I} 

To obtain the distance modulus of Leo~I we adopt the calibration of
$M^{TRGB}_I$ as a function of the {\em global metallicity} ($[M/H]$) recently
provided by \citet{tip2} \begin{equation}
M^{TRGB}_I=0.258[M/H]^2+0.676[M/H]-3.629 ~~~\pm 0.12. \end{equation} The global
metallicity includes   the contribution of Iron and the $\alpha$-elements (O,
Mg, Ti, Si, etc.),  hence it is a more suitable indicator of the global metal
content to relate with the  observed properties of stars and stellar
populations and for comparisons with theoretical models \cite[see][for
discussion and references]{scs93,f99,tip2}. The parameter is defined as
\begin{equation} [M/H]=[Fe/H]+log(0.638\times10^{[\alpha/Fe]}+0.362)
\end{equation} \citep{scs93}. We also adopt $E(B-V)=0.01\pm 0.01$, according to
\citet{mateo} and   $A_I=1.76E(B-V)$, according to \citet{dean}. 

At the mean metallicity  $[M/H]=-1.2$ 
(as derived in Sect.~3.3, below), the resulting distance modulus 
is $(m-M)_0=22.02$.

This estimate is affected by the combination of
uncertainties coming from different sources, i.e. the estimate of apparent
magnitude of the TRGB, the calibrating relations, the reddening and the assumed
global metallicity.  To properly account for all these uncertainty factors we
recur to a Monte Carlo simulation. We compute the final true distance modulus
(and its uncertainty) as the mean (and standard deviation) of 100000 random
realizations of the $(m-M)_0$ estimate obtained by allowing all the relevant
ingredients to vary within their error bars. In particular:

\begin{itemize}

\item The {\bf apparent TRGB magnitude} was randomly extracted from two
gaussian distributions with mean equal to the observed value. 50000 positive
deviations were extracted from a gaussian with $\sigma=0.05$ and 50000 negative
deviations were extracted from a gaussian with $\sigma=0.03$. To account for
the uncertainty in the absolute photometric calibration we add to each
realization of $I^{TRGB}$ a component extracted from a gaussian distribution
having zero mean and $\sigma=0.02$ mag.

\item We added to {\bf the absolute TRGB magnitude} (from the calibrating 
      relation by B04) a random component extracted from a gaussian
      distribution having zero mean and standard deviation equal to the
      uncertainty of the zero point of the calibrations as estimated by B04,
      i.e. $\sigma=0.12$.

\item The {\bf reddening} values were extracted from a gaussian distribution
      with mean $E(B-V)=0.01$ and $\sigma_{E(B-V)}=0.02$ to account also for
      slightly different reddening estimates found in literature  \cite[for
      instance the reddening maps by][suggests E(B-V)=0.03]{cobe}. Negative
      values of the reddening are excluded from the simulation.

\item The adopted {\bf metallicity} values were extracted from a gaussian 
      distribution with mean $[M/H]=-1.2$ and $\sigma=0.2$. 

\end{itemize}

For each simulated point $(m-M)_0$ was computed from the randomly extracted
$I^{TRGB}$, $M_I^{TRGB}$, $E(B-V)$ and $[M/H]$. 
Finally, the mean $<(m-M)_0>$  and
$\sigma$ of the 100000 realizations was obtained: $(m-M)_0=22.02\pm 0.13$, 
corresponding to a heliocentric distance $D=254^{+16}_{-19}$ Kpc. This result
is in excellent agreement with the value tabulated in the recent review by
\citet{mateo}, but the uncertainty is reduced by a factor $\sim 2$ by the
present analysis \cite[see also][who finds $(m-M)_0 = 22.05 \pm 0.10$
(internal) $\pm 0.18$ (systematic)]{mendez}. It is interesting to note that our
distance modulus is in excellent agreement with the recent estimate by
\citet{held_RR} based on the mean V magnitude of 47  RR(ab) Lyrae variables
discovered by these authors.


\subsection{Metallicity Distributions}

In Fig.~5 the observed RGB of  Leo I is compared with the template
ridge lines taken from the set adopted by \citet{m31}.  
It is immediately clear that the bulk of Leo~I RGB
stars is enclosed  within the ridge lines of NGC~6341 ($[Fe/H]_{CG}=-2.16$;
$[Fe/H]_{ZW}=-2.24$; $[M/H]=-1.95$) and of  NGC~5904 ($[Fe/H]_{CG}=-1.11$;
$[Fe/H]_{ZW}=-1.40$; $[M/H]=-0.90$). The average Color-Magnitude distribution
is well represented by the ridge line of NGC~6205 ($[Fe/H]_{CG}=-1.39$;
$[Fe/H]_{ZW}=-1.65$; $[M/H]=-1.18$). Consequently, we adopt $[M/H]=-1.2$ as 
the mean metallicity of Leo I. The age difference between the bulk of the Leo I
population and the adopted templates \citep{carme} may slightly affect this
estimate. However this is not a reason of serious concern for the present
application, since the dependence of $M_I^{TRGB}$ on metallicity is very weak
for $[M/H]\le 0.5$ \cite[see also][ and references therein]{mf98}.

\begin{figure} 
\includegraphics[width=84mm]{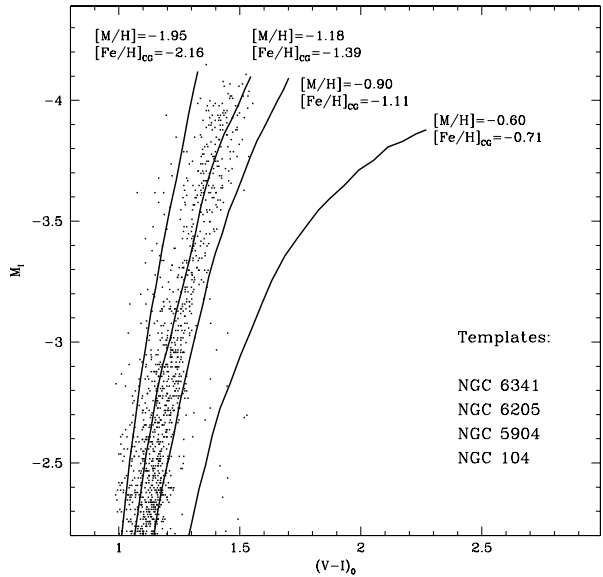}  
\caption{Comparison of
the observed RGB of Leo~I (assuming $(m-M)_0=22.02$ and $E(B-V)=0.01$) with the
ridge lines of the templates adopted by \citep{m31}.  
The bulk of Leo~I RGB stars is enclosed between the ridge lines of NGC~6341 and
NGC~5904.} \end{figure}

 Our result is broadly consistent with all previous photometric estimates of
the mean metallicity of Leo~I, all in the ZW scale. In terms of
mean/median metallicity our estimates lie between the extremes provided by
\citet{reid} ($[Fe/H]_{ZW}=-1.0\pm 0.3$) and \citet{lee_leo} 
($[Fe/H]_{ZW}=-2.1\pm 0.1$), in good agreement with the low-resolution spectral
analysis by \citet{suntz} and its revision reported by \citet{lee_leo} 
($[Fe/H]_{ZW}\sim -1.8$, from 2 red giants) and, finally, in excellent
agreement with \citet{gambu} who finds $[Fe/H]_{ZW}=-1.6\pm 0.4$

\citet{shet} reported the first detailed abundance analysis of two
Leo~I red giants, based on high resolution spectroscopy (hence, presumably, in
a scale equivalent to the CG one). They find $[Fe/H]=-1.5$ and $[Fe/H]=-1.1$
for the two considered stars.  
It is interesting to note that the same authors provides
also an estimate of the $[\alpha/Fe]$ ratio for these stars, i.e.
$[\alpha/Fe]\simeq +0.5$ and $[\alpha/Fe]\simeq 0.0$, respectively. From these
values, using Eq.~2, we find that both stars have $[M/H]=-1.1$, in excellent
agreement with our estimate. 

\section{Conclusions}

We have provided a clean and accurate detection of the I magnitude of the TRGB
of Leo~I from two independent datasets, tied to the same (well checked)
photometric calibration. Adopting the median metallicity we derived from  the
same data by comparison with templates RGB ridge lines, and the calibration of
$M^{TRGB}_I$ as a function of the global metallicity ($[M/H]$) provided by
\citet{tip2} we have obtained a new estimate of the distance modulus of Leo~I,
$(m-M)_0=22.02 \pm 0.13$, corresponding to a  distance $D=254^{+16}_{-19}$ Kpc.

The effects of all the possible sources of uncertainty have been taken into
account by means of Monte Carlo simulations.  The obtained distance modulus is
in  good agreement with the most recent distance estimates for this galaxy but
it reduces the uncertainty by a factor $\sim 2$, with respect to previous
applications of the TRGB method.

\section*{Acknowledgments}

This research is partially supported by the italian {Ministero 
dell'Universit\'a e della Ricerca Scientifica} (MURST) through the COFIN grant
p.  2002028935-001, assigned to the project  {\em Distance and stellar
populations in the galaxies of the Local Group}. 
Part of the data analysis has
been performed using software developed by P. Montegriffo at the INAF -
Osservatorio Astronomico di Bologna. 
This research has made use of NASA's
Astrophysics Data System Abstract Service.

\label{lastpage}

\end{document}